\documentclass[a4paper, 12pt]{amsart}
\usepackage{graphicx,amscd}
\newtheorem{theorem}{Theorem}[section]

\title{ Quantum E-Cheques}
\author{Do Ngoc Diep${}^{1,3}$}
\address{${}^1$ Instittute of Mathematics, Vietnam Academy of Science and Technology, 18 Hoang Quoc Viet road, 10307 Hanoi, Vietnam}
\email{dndiep@math.ac.vn}
\author{Nguyen Van Minh${}^2$}
\address{${}^2$ Department of Mathematics, Thuong Tin High School, Tran Phu Road, Thuong Tin town, Thuong Tin District, Hanoi Vietnam.}
\email{nvminh07@gmail.com}
\address{${}^3$ Institute of Mathematics and Applied Sciences, Thang Long University, Nghiem Xuan Yem road, Hoang Mai district, Ha Noi, Vietnam} 
\begin{document}
\maketitle
\begin{abstract}
We analyze the procedure providing quantum cheques of S. R. Moulick and P. K. Panigrahi \cite{moulick-panigrahi} to produce quantum e-cheques, based on multiparty quantum telecommunication between custumer and cooperated branches of  bank.
\end{abstract}
\subjclass{\sl AMS Mathematics Subject Classification:} 15A06; 15A99

\keywords{\sl Keywords and Terms:} quantum secret sharing scheme; quantum multivariate interpolation, quantum cheques, quantum e-cheques

\section{Introduction}
The problem of providing a quantum code of  classical cheques is a central problem of the so called quantum money problem. The question is to provide a scheme of quantum code in such a way that it should be similar to the classical ones but with absolute high secrecy. In the work \cite{moulick-panigrahi}, the authors gave an adequate survey of development of the problem and constructed a scheme for quantum cheques.
The scheme is covered the classical version of cheques: The quantum cheque will use the schemes of the form $\prod = (Gen, Sign, Vrfy)$. 

Some customer Alice and bank make initialation by the
\\
\textbf{Gen scheme:} namely  Alice came to some bank branch to open an account with \textsl{secret key} as a binary $L$-digit number $k\in \{0,1\}^L$ to provide an electronique \textsl{signature} in the future, by using some \textsl{secret key generation scheme} for Alice and bank. 
The bank later gives her a \textsl{cheque book serial number} $s$. For secrecy, Alice produces some  \textsl{public key} $pk$ and store a  \textsl{secret key} $sk$. 
The bank produces 3 entangled qubits in GHZ  states
$$|\phi^{(i}\rangle_{GHZ} = \frac{1}{\sqrt{2}}\left( |0^{(i)}\rangle_{A_1} |0^{(i)}\rangle_{A_2}|0^{(i)}\rangle_{B}+|1^{(i)}\rangle_{A_1} |1^{(i)}\rangle_{A_2}|1^{(i)}\rangle_{B}\right), 1 \leq i \leq n$$
and send two of them, namely $|\phi\rangle_{A_1}$ and $|\phi\rangle_{A_2}$ to Alice. Therefore Alice holds $(id, pk, sk, k, s, \{|\phi^{(i)}\rangle_{A_1}, \phi^{(i)}\rangle_{A_2}\})$ and the bank branch holds $(id, pk, sk, k, s, \{|\phi^{(i)}\rangle_{B}\})$. 

The next step is the 

\textbf{Sign scheme:}.
Alice chooses a random number $r$ with use a random number generation procedure $r\leftarrow U_{\{0,1\}^L}$, 
 a numeration $i= 1,\dots, n$ of orthogonal base $|\phi^{(i)}\rangle$ and certainly an amount $M$ she likes to make some transaction with bank (debit or credit), and then evaluate the  one-way funtion $f: \{0,1\}^* \times |0\rangle \to |\psi^{(i)}\rangle$ at the concatennation $x||y$ of the data as $k||id|| r ||M||i$ to provide a state $\psi^{(i)} = \alpha_i |0\rangle + \beta_i|1\rangle$. Alice encodes the data $|\psi^{(i)}\rangle$ with the $|\phi^{(i)}\rangle_{A_1}$, making them entangled and measuring the Bell states:
 $$|\phi^\pm\rangle = \frac{1}{\sqrt{2}}\left(|00\rangle \pm |11\rangle \right),\quad  |\psi^\pm\rangle = \frac{1}{\sqrt{2}}\left(|10\rangle \pm |01\rangle \right)$$. The system is in the states of form
 $|\phi^{(i)}\rangle = |\psi^{(i)} \rangle \otimes |\phi\rangle_{GHZ} = $ $$\frac{1}{2} \left\{|\phi^{+}\rangle_{A_1}(\alpha_i |00\rangle_{A_2B} + \beta_i|11\rangle_{A_2B} ) +
 |\phi^{-}\rangle_{A_1}(\alpha_i |00\rangle_{A_2B} - \beta_i|11\rangle_{A_2B} )\right.$$
$$\left. +|\psi^{+}\rangle_{A_1}(\alpha_i |00\rangle_{A_2B} + \beta_i|11\rangle_{A_2B} )+|\psi^{-}\rangle_{A_1}(\alpha_i |00\rangle_{A_2B} - \beta_i|11\rangle_{A_2B} )
  \right\}$$
  Then Alice performs Pauli transforms 
  $$|\phi^+\rangle \to I=\begin{pmatrix} 1 & 0\\ 0 & 1\end{pmatrix}, \quad |\phi^-\rangle \to \sigma_Z = \begin{pmatrix} 1 & 0\\ 0 & -1\end{pmatrix}$$ 
  $$|\psi^+\rangle \to \sigma_X=\begin{pmatrix} 0 & 1\\ 1 & 0\end{pmatrix}, \quad |\psi^-\rangle \to \sigma_Y = \begin{pmatrix} 0 & -i\\ i & 0\end{pmatrix}$$ 
 and make correction to $|\phi^{(i)}\rangle_{A_2}$
  Alice makes signature by using the procedure $sign_{pk}(s)$ and produces the quantum cheque $\chi = (id,s, r,\sigma,M,\{|\phi^{(i)}_{A_2}\})$ then publicly send through Abby to an arbitrary of the valid branches of the bank.
 
 The final step is the verification\\
 \textbf{Vrfy scheme:} A valid bank branch after received the cheque, informs to the main branch in order to check the signature $Vfry(\sigma, s)$. For this one uses namely the well-known Fredkin gate (\cite{moulick-panigrahi}, Picture 1).
 If the $(id, s)$ or $\sigma$ is invalid, the bank destroy the cheque, otherwise the bank continue the measurement in Hadamard basis $|\phi\rangle_B$. If the result is $|\phi^+\rangle$ or $|\phi^-\rangle$ the main branch communicates to the acting banch to continue. The acting branch perform transformation $|\phi^+\rangle \to I$ and $|\phi^-\rangle \to \sigma_Z$. The bank accepts the cheque if it passes the swap test then destroy it.
 
The schemes are summarized as in the Figure 1 of \cite{moulick-panigrahi}:
\begin{figure}[!ht] 
			\centering
				\includegraphics[scale=0.4]{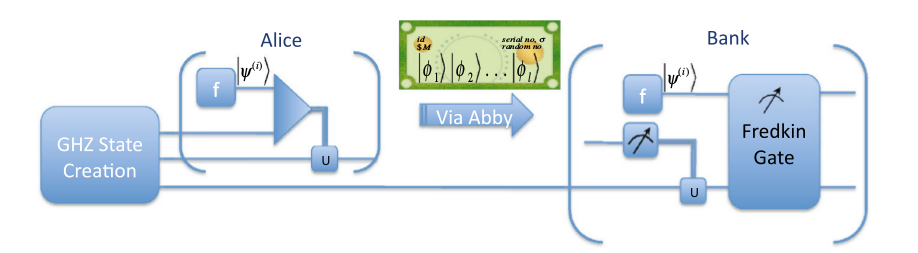}
                    \caption{quantum cheques transfering code}
\end{figure} 

We remark that the quantum cheque is produced and used quite similar to the classical one. We propose therefore to use the multipartite quantum key distribution to make quantum cheques become quantum e-cheques of high secrecy \cite{diep1}. Our main result is Theorem \ref{thm1} stating that a code for quantum e-cheques can be provided with high secrecy by the multipartite public key distribution of quantum share.  The feature of our approach is that (i) Alice does not need to go to a bank branch to do a transaction, but divides her data to a disjoint union of parts and connects with acting branches to send to each one part of her data, (ii) the bank can record the Alice's data only if all acting branches cooperate together and therefore (iii) they defense the origin of data and Alice prevents some dishonest branches to change the data. 

We separately consider the problem of e-cheque transfering in the situation of absolute secured channel in order to point out the main idea of e-chequering.  
The more complicated problem of e-cheque transfering in presence of eavesdroppers in a nonserured channel or dishonest participants will be separately considered in a subsequent paper. 

We thank Prof. S. R. Moulick and  Prof. P. K. Panigrahi for carefully reading the first draft of this paper, for valuable comments and for reminding the authors about mediator.

The paper is devoted to this construction in the next Section 2 and finishes with some conclusion.

\section{Quantum e-cheques as multiparty quantum secrete sharing}

Consider the following \textit{modified problem} for the situation when Alice does not send the quantum cheques via Abby, but could online connect with acting branches of a bank. To  prevent the fact that some distrusted branches could change the cheque. The bank could discover the informations from the quantum cheques only if all acting branches cooperate togheter and in that case the other branches prevent the some untrusted branches to change the contents of the cheque. The quantum cheques in that case are what we call \textbf{e-cheques}. 

Solution to this problem is the following scheme of code.

After the first step \textbf{Gen scheme}, in the second step\\ \textbf{Sign Scheme} one keeps the same as in the previous section, only now, Alice divides the provided concatened information  $k||id|| r ||M||i$  into $n$ parts, $D^{(i)}_1,\dots,D^{(i)}_n$, where $n$ is an appropriate number of branches in action. Then she produces the corresponding states by using a one-way function $f$ to have $f(D^{(i)}_j) =\psi^{(i)}_j$, for all $j = 1,\dots, n$. Following the multiparty secret sharing, when the branches cooperate together and inform to the main branch, one discover the states $|\phi^{(i)}_j\rangle$

\begin{theorem} \label{thm1} The quantum cheques could be with higher secrecy electronically transfered from Alice to the acting bank branches by a code of multiparty quantum telecommunication problem of secret sharing with quantum public key distribution.
\end{theorem}
\textit{Proof.} The theorem is proved by the 
 following 
 
 \textbf{Procedure}, which is similar to the one  in the 3 persons case by Cabello \cite{cabello}, following which the system states are changing as follows.
 $$\CD |\psi_i\rangle @>>> |\psi_{ii}\rangle @>>> |\psi_{iii}\rangle @>>> |\psi_{iv}\rangle \endCD$$ Let us consider it in more details.
  
\textit{Transfer Step 1. Initialization of 3n qubits.} For a fixed $i$, Alice uses $n+2$ qubits, named: $1,2,3, D_1=D_1^{(i)},\dots,  D_{n-1}=D_{n-1}^{(i)}$: qubits 1 and 2 are entangled in Bell state, qubits $3, D_1, \dots, D_{n-1}$ are entangled in GHZ state with $n-1$ acting bank branches: Branch 1, Branch 2, ...., Branch n-1, each has 2 entangled qubits $i+3, C_i,i=1\dots,n-1$ namely in null state. Alice produces a Bell state measurement on qubit 1 and 2 and a Fourier measurement \frame{$F_n$} on $n$ qubits $3, D_1,\dots, D_{n-1}$. Each of acting branch makes a Bell state Fourier measurement  \frame{$F_2$} of entangled $i+3,C_i, i=1\dots n-1$. 
At the end of this step 1, the system is in the state
$$|\psi_i\rangle = |0\dots 0\rangle_{3D_1\dots D_{n-1}} \otimes |00\rangle_{12} \otimes |00\rangle_{4C_1}\otimes\dots \otimes |00\rangle_{n+2,C_{n-1}} $$

\textit{Transfer Step 2. Entangled Bell-state measurements.}
Alice sends each qubit $D_i$  of her GHZ state out to
each $i^{th}$ acting bank branch of the other $n-1$ branches. 
The system is in the state
$$|\psi_{ii}\rangle = |AP\rangle_{3D_1\dots D_{n-1}} \otimes |BP\rangle_{1C_1} \otimes|CP\rangle_{2C_2} \otimes\dots \otimes |NP\rangle_{{n+2},C_{n-1}} $$

\textit{Transfer Step 3. Secret Bell-state measurement.}
Next, Alice and each user  performs a Bell-state Fourier measurement  \frame{$F_2$} on the received qubit and one of their qubits.
After these measurements the state of the system
becomes
$$|\psi_{iii}\rangle = |AP\rangle_{3D_1\dots D_{n-1}} \otimes |AS\rangle_{2,3}\otimes |BS\rangle_{4,D_1} \otimes \dots \otimes |NS\rangle_{n+2,D_{n-1}},$$
where $|AP\rangle$ is $n$-qubit GHZ state of the standard orthonormal basis.

\textit{Transfer Step 4. Secret sharing.} 
The $n-1$ acting branches send a qubit (the one they have not used) to Alice, and she performs a Fourier measurement  \frame{$F_n$} to discriminate between the $2^n$ GHZ states, and publicly announces the result $|AP\rangle_{1C_1\dots C_{n-1}}$.
After these measurements the state of the system
becomes
$$|\psi_{iv}\rangle = |AP\rangle_{1C_1\dots C_{n-1}} \otimes |AS\rangle_{2,3}\otimes |BS\rangle_{4,D_1} \otimes \dots \otimes |NS\rangle_{n+2,D_{n-1}},$$
 The result AP, and the result of their own secret measurement allow each legitimate acting branch to infer the first bit of Alice’s secret result $AS$. To find out the second bit of Alice's secret $AS$, all users (except Alice) must cooperate. 
 
 In case $n=3$ there is an illustration of Cabello \cite{cabello} as in Figure 2.
 \begin{figure}[!ht] 
			\centering
				\includegraphics[scale=0.4]{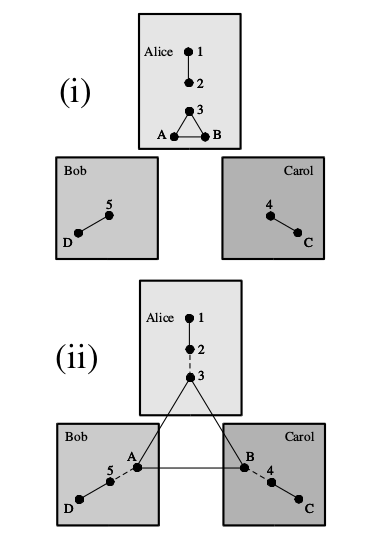}
                             \includegraphics[scale=0.4]{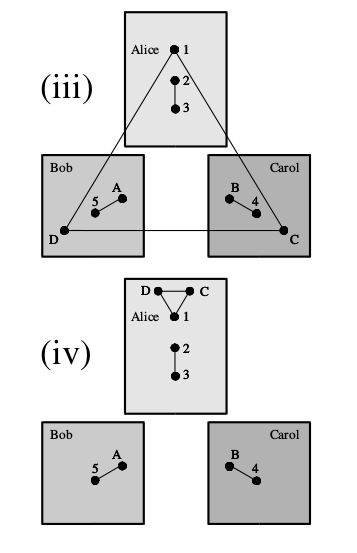}
                    \caption{Quantum cheques transfer code with $n=3$}
\end{figure} 
 The proof therefore is achieved. \hfill$\Box$
  
The 4 steps scheme of public key secret sharing distribution can be generalized to the case of two levels groupped secret sharing as illustrated in the work of A. Jaffe, Z.-W. Liu, and A. Wozniakowsk\cite{jaffe} in Figure 3:

\begin{figure}[!ht] 
			\centering
				\includegraphics[scale=0.4]{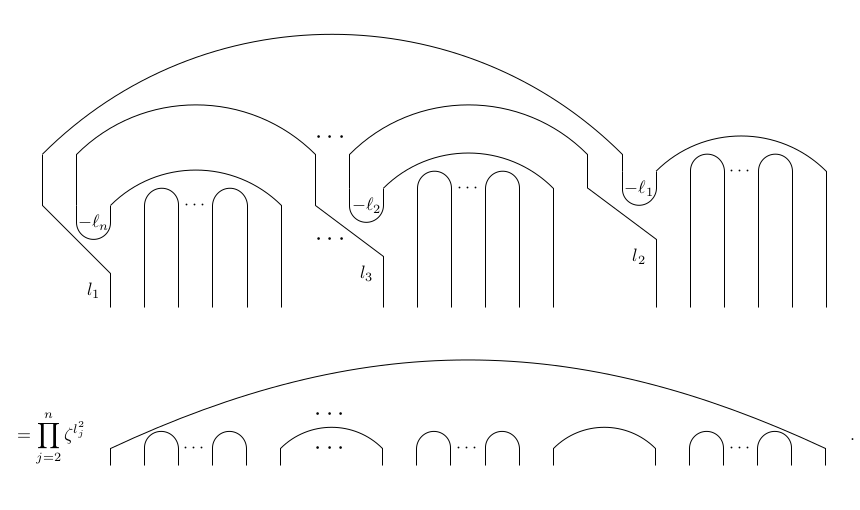}\quad
                             \includegraphics[scale=0.4]{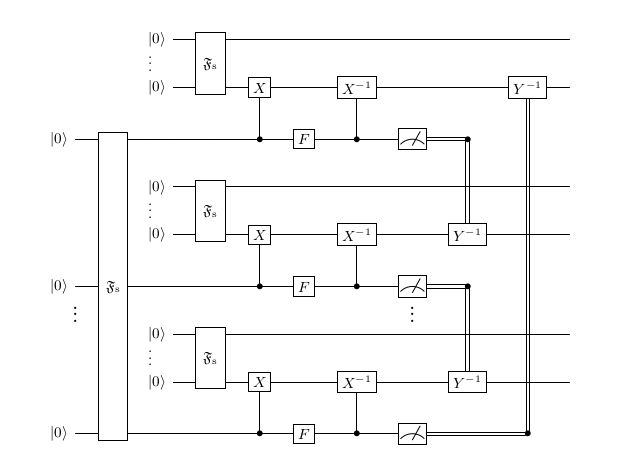}
                    \caption{n-partitie-sharing problem and corresponding BVK code}
\end{figure} 

After discovered the e-cheque, bank continue to procede the same procedure \textbf{Vrfy Scheme}  as in the quantum cheques scheme above to verify the validity of the e-cheque and accept of destroy it.

\section{Conclusion}
We show that the quantum cheques can be electronically transfered with higher secrecy by a code of multiparty quantum telecommunication problem of secret sharing with quantum public key distribution. The problem of transfering e-cheques in nonsecured channel is separately considered in a subsequent paper.

\end{document}